\documentclass[english,pre,twocolumn,amsmath,amssymb,floatfix,showpacs]{revtex4-1}
\usepackage[utf8x]{inputenc}
\usepackage{bm}
\usepackage{color}
\usepackage{graphicx,epsfig}
\newcommand{\aseq}{\simeq}

\begin{document}

\title{Comment on ``Nonlocal quartic interactions and universality classes in perovskite manganites''} 
\author{H.~W. Diehl}
\affiliation{Fakult\"at f\"ur Physik, Universit\"at Duisburg-Essen, D-47048 Duisburg, Germany}

\pacs{05.70.Jk,05.10.Cc,75.40.Cx}
\date{\today}

\begin{abstract}
In a recent paper [Phys.\ Rev.\ E \textbf{92}, 012123 (2015)]  a modified $d$-dimensional $\Phi^4$ model was investigated that differs from the standard one in that the $\Phi^4$ term was replaced by a nonlocal one with a potential $u(\bm{x}-\bm{x}')$ that depends on a parameter $\sigma$ and decays exponentially as $|\bm{x}-\bm{x}'|\to\infty$ on a scale $|m|^{-1}<\infty$.  The authors claim  the upper critical dimension of this model to be $d_\sigma=4+2\sigma$. Performing a one-loop calculation they arrive at expansions in powers of $\epsilon_\sigma=d_\sigma-d$ for critical exponents such as $\eta$ and related ones to $O(\epsilon_\sigma)$  whose $O(\epsilon_\sigma)$ coefficients depend on $\sigma$ and the ratio $w=m^2/\Lambda^2$, where $\Lambda$ is the uv cutoff. It is shown that these claims are unfounded and based on misjudgments and an ill-conceived renormalization group (RG) calculation. \end{abstract}

\maketitle

The authors of \cite{SDN15} investigated an $n$-component $\Phi^4$ model in $d$ dimensions  with a nonlocal quartic term  whose Hamiltonian is given by
\begin{eqnarray}\label{eq:H}
\mathcal{H}&=&\int d^dx\bigg[\frac{c_0}{2}|\nabla\Phi(\bm{x})|^2+\frac{r_0}{2}\Phi^2(\bm{x})\nonumber \\ &&\strut +\int d^dx'\Phi^2(\bm{x})\,u(\bm{x}-\bm{x}')\,\Phi^2(\bm{x}')\bigg],
\end{eqnarray}
where the potential is specified by its Fourier transform
\begin{equation}\label{eq:utilde}
\tilde{u}(\bm{k})=\frac{\lambda_0}{[k^2+m^2]^\sigma}
\end{equation}
with a ``screening parameter'' $m^2>0$. From Eq.~\eqref{eq:utilde} one can read off   that $u(\bm{x})$ falls off exponentially on the scale $\xi_m=|m|^{-1}$ as $|\bm{x}|\to \infty$ \footnote{The Fourier back-transform of $\tilde{u}(\bm{k})$ (calculated by extending the $\bm{k}$-integration to $\mathbb{R}^d$) gives $u(\bm{x})=%
\pi^{-d/2}2^{1-d/2-\sigma }( |m|/x)^{d/2-\sigma }K_{d/2-\sigma }(|m| x)/\Gamma (\sigma )$. The asymptotic form of the potential is $u(\bm{x})\aseq 2^{(1-d)/2-\sigma}\pi^{(1-d)/2}(|m|/x)^{(d-1)/2-\sigma }x^{-1}e^{-|m| x}
 /\Gamma(\sigma )$.    }. Hence  it  is short-ranged on the scale of the correlation length $\xi(T)$ sufficiently close to the critical temperature $T_c$. The reason is that $\xi_m$ is  temperature-independent or  has at most a weak temperature dependence when the interaction is considered as an effective interaction obtained upon coarse graining  to a length scale $2\pi/\Lambda$ ($>a$) on which a continuum approximation is acceptable. Therefore $\xi_m$ remains bounded whereas $\xi(T)\to\infty$ as  $T\to T_c$, so that $\lim_{T\to T_c}\xi_m/\xi\to 0$ \footnote{For the sake of simplicity and clarity, my reasoning here is tailor-made for the $n=1$ case. When $n>1$, so that one is considering $O(n)$ vector models, the given arguments apply to the disordered phase (which is sufficient for the my conclusions concerning the universality class). In a proper discussion of the low-temperature ordered phase ($n>1$) one would have to make the familiar distinction between longitudinal and transverse correlation lengths and cope with the spontaneously broken symmetry, of course. But this would not change my conclusions about the universality class of the mode.}. What matters for the asymptotic critical behavior is just whether or not the interaction is short-ranged on the scale of $\xi(T)$ as $T\to T_c$. This is asymptotically always the case \emph{even if  $\xi_m$ is much larger than the lattice constant $a$, i.e., if $u(\bm{x})$ is long-ranged on the scale of $a$}. Therefore the model must belong to the standard universality class of the $d$-dimensional $n$-vector model with short-range interactions represented by the usual $\Phi^4$ model with a local quartic term.
 
The authors' results strongly disagree with this. Performing a one-loop calculation, they conclude that the upper critical dimension of the model is $d_\sigma=4+2\sigma$, and they compute some of  the critical exponents of the model by an expansion in $\epsilon_\sigma=d_\sigma-d$ to first order. The $O(\epsilon_\sigma)$ coefficients of their series expansions depend on $\sigma$ and $m^2/\Lambda^2$. If this were true, the critical exponents themselves would have these properties, and hence depend on microscopic details such as $m^2/\Lambda^2$ and be nonuniversal. Obvious serious discrepancies with the modern theory of critical behavior (see, e.g., \cite{Fis98,Ami89} and references therein) would exist.
Aside from a gross violation of universality, one would have the following problems: For $\sigma>0$, the  asymptotic critical behavior of the model in $4<d<d_\sigma$ dimensions would be inconsistent  with mean-field theory. Likewise, irrespective of whether  $\sigma>0$ or $\le 0$, the critical behavior of the model would differ in $d_*<d<4$ dimensions from that of the standard $\Phi^4$ model., where $d_*$ is the lower critical dimension ($=1$ or $2$ for $n=1$ and $n>1$, respectively).  However, the authors' claims and their consequences are unfounded: Their results are based on an ill-conceived renormalization-group (RG) analysis, and they are therefore incorrect.

To see what goes wrong with their analysis \footnote{The constant $c_0$ can be absorbed in the amplitude of $\bm{\Phi}$ and hence be set to $1$.}, note that their one-loop calculation of the four-point vertex function yields a modified quartic interaction that does not comply with the original one, i.e., with $\tilde{u}(\bm{k})$. This is, of course, expected. The authors then drop the contributions that do not comply with the original nonlocal form of the quartic interaction, arguing that they must be irrelevant. This is incorrect: All three of their graphs (a), (b), and (c) contribute to the transformed coupling constant $u_0'$ associated with the local part of the $\Phi^4$ interaction one derives from $\Gamma^{(4)}(\bm{k}_1,\ldots,\bm{k}_4)\big|_{\{\bm{k}_j=\bm{0}\}}$, as can be seen upon substitution of the expansion $\tilde{u}(\bm{k})=u_0 +O(k^2)$ into the corresponding integrals with $u_0=\lambda_0 m^{-2\sigma}$. Using this expansion is unproblematic since  $\tilde{u}_0(\bm{k})$ is analytical in $k^2$ as long as $m^2>0$. In fact, the sum of the contributions from the graphs (a), (b), and (c) yield the usual result for $u_0'$. Furthermore,  the contributions arising from the $k^2$-dependent parts of $\tilde{u}(\bm{k})$ can be dropped because they are irrelevant. This follows in the usual way  via power counting \cite{Ami89}: The interaction constant $u_0=\lambda_0m^{-2\sigma}$ has the familiar momentum dimension $\epsilon=4-d$.  The coupling constant $\lambda_0m^{-2(1+\sigma)}$ of the momentum-dependent quartic term arising from the contribution $\propto k^2$ to $\tilde{u}(\bm{k})$ has the momentum dimension $\epsilon-2$. Thus it is irrelevant at the Gaussian fixed point whenever $0<\epsilon<2$. The corrections to the leading asymptotic behavior and the associated correction-to-scaling exponents can be --- and have been --- computed within the framework of the usual $\epsilon$ expansion. To do this one must study the insertions of  all composite operators  of the same or lower momentum dimensions (see, e.g., Chapter 10 of \cite{Ami89} or \cite{KW94}). Upon dropping the irrelevant $\bm{k}$-dependent parts of $\tilde{u}(\bm{k})$, one recovers the standard $\Phi^4$ model with a local $\Phi^4(\bm{x})$ term, which therefore represents the universality class of the model~\eqref{eq:H}. 

For readers with a background in critical phenomena  and  the basics of the RG it will be clear that the analyticity of $\tilde{u}(\bm{k})$ at $k=0$  ensures that the used expansion in $\bm{k}$ is both convenient and safe in the $m>0$ case.  They will also know that such an expansion in k is precisely what one does in a standard momentum-shell RG approach \cite{WK74} to  identify both the transformed interaction constants into which the original  ones transform under RG transformations as well as those generated under the RG. Readers who lack this familiarity with these scientific areas might wonder whether different conclusions could be obtained when $m>0$ if the expansion of $\tilde{u}(\bm{k})$ in $\bm{k}$ is avoided. The answer to this question is a clear no. To see this let us introduce the dimensionless coupling constant  $g=\mu^{-\epsilon} \tilde{u}(\bm{0})=\mu^{-\epsilon}\lambda_0m^{-2\sigma}$, where $\mu$ is an arbitrary momentum scale, and decompose $\tilde{u}(\bm{k})=\mu^\epsilon g+\tilde{u}_1(\bm{k}))$ into its $\bm{k}=0$ part and a remainder $\tilde{u}_1(\bm{k})=\lambda_0m^{-2\sigma}[(1+k^2/m^2)^{-2\sigma}-1] $. In the absence of the latter, there is a well-defined RG in $d=4-\epsilon<4$ dimensions which implies that $g$ transforms under a change of scale $\mu\to\mu \ell$ into a running coupling constant $g(\ell) $ that approaches a fixed-point value $g^*=O(\epsilon)$ in the infrared limit $\ell\to 0$. The scaling dimension $\Delta_\varepsilon$ of the energy-density operator $\bm{\Phi}^2(\bm{x})$ at this fixed point is $d-1/\nu$, where $\nu=1/2+O(\epsilon)$ is the correlation-length exponent; i.e., $\bm{\Phi^2}(\bm{x}\ell)\sim \ell^{-\Delta_\varepsilon}\bm{\Phi}^2(\bm{x})$. Upon including $\tilde{u}_1(\bm{k})$ to obtain the full interaction $\tilde{u}(\bm{k})$, we can conclude that $\tilde{u}_1(\bm{k})$ transforms into $\tilde{u}_1(\bm{k},\ell)=\ell^{d-2/\nu}\tilde{u}_1(\ell\bm{k})$. But $\tilde{u}_1(\bm{k})=O(k^2)$ as $k\to 0$. Hence $\tilde{u}_1(\bm{k},\ell)$ varies $\sim\ell^{2+d-2/\nu}$ as $\ell\to 0$, where the exponent $2+d-2/\nu=2+O(\epsilon)>0$. Thus  $\tilde{u}_1(\bm{k})$ is strongly irrelevant and flows to zero under the RG in the infrared limit $\ell\to 0$. 

As a consequence, one arrives at the following mathematically exact statement: The critical exponents of the $m>0$ model studied in Ref.~\cite{SDN15} have a well-defined $\epsilon$ expansion, where the associated series-expansion coefficients of any order in $\epsilon$ have exactly the same values as for the standard $\Phi^4$ model with a local $\Phi^4(\bm{x})$ interaction.

Analogous considerations for $d>4$ imply that both $g$ and $\tilde{u}_1(\bm{k})$ are irrelevant. One concludes that mean-field theory holds for $d>4$, where  one must of course take into account  that $g$ is a dangerous irrelevant variable \cite{Ami89} which hence cannot be set to its fixed-point value $g^*=0$ in quantities such as the free energy or $\langle\Phi(\bm{x})\rangle$ below $T_c$.

It may also be mentioned that this model can  be described equivalently by a two-field model of the form used in the definition of the familiar stochastic model C \cite{HH77}, namely a model with the Hamiltonian
\begin{eqnarray}\label{eq:Ham}
\mathcal{H}'&=&\int d^3x\bigg[\frac{c_0}{2}\bm{\Phi}^2+\frac{1}{2}\psi\,(m^2-\nabla^2)^{\sigma}\psi\nonumber\\\ &&\strut +\gamma_0\,\psi\,\bm{\Phi}^2+v_0|\bm{\Phi}|^4\bigg],
\end{eqnarray}
where $\psi(\bm{x})$ is a one-component field. The interaction term now is local. The Hamiltonian~\eqref{eq:H} is easily recovered  (up to field-independent terms) from $\mathcal{H}'$ with the choices $\gamma=i\sqrt{2\lambda_0}$ and $v_0=0$ by integrating out $\psi$. We have included a local $|\bm{\Phi}|^4$ in $
\mathcal{H}'$ since it would generically be present anyway both in $\mathcal{H}$ and $\mathcal{H}'$ \footnote{In the case of model C, the interaction constant $\gamma_0$ would be taken to be real-valued. Choosing the coefficient of the $\psi^2/2$-term to be a constant $C_\psi=m^{2\sigma}$, one arrives at a standard $\Phi^4$ model with a coupling constant $u_0=v_0-\gamma^2/(2C_\psi)$ by integrating out $\psi$.}.

In a momentum-shell RG  scheme, one would have to choose the field-rescaling factors $b_\Phi$ and $b_\psi$ such that the coefficient $c_0$ of the $|\nabla\Phi|^2/2$ term and the coefficient $m^{2\sigma}$ of the $\psi^2/2$ term are preserved. The again, familiar considerations show that the terms quadratic in $\psi$ that involve $\nabla^2$ are irrelevant. The same conclusion can also be reached by simple power counting. Thus one recovers the above-mentioned findings: The model belongs to the universality class of the standard $\Phi^4$ model and its critical exponents have the familiar dimensionality expansions in $\epsilon=4-d$ for $d<4$. 

These conclusions can be corroborated by solving the model~\eqref{eq:Ham} with ${m^2> 0}$  exactly in the limit $n\to\infty$. For $n=\infty$, the critical exponents must take their usual ${n=\infty}$ values, namely, $\nu_\infty=\gamma_\infty/2=(d-2)^{-1}$, $\beta_\infty=\nu_\infty(d-2+\eta_\infty)/2=1/2$, $\alpha_\infty=2-d\nu_\infty=(d-4)/(d-2)$ and $\delta_\infty=(d+2)/(d-2)$ for $2<d<4$, and the mean-field values $\nu_{\text{MF}}=\gamma_{\text{MF}}/2=\beta_{\text{MF}}=1/2$, $\eta_{\text{MF}}=\alpha_{\text{MF}}=0$ and $\delta_{\text{MF}}=3$ for $d>4$.  To see this note that  the self-consistent equation for the inverse susceptibility $r=\tilde{\Gamma}^{(2)}({\bm{k}=\bm{0}})$ \emph{becomes identical} to its analog for the standard $\Phi^4$ theory with a local quartic interaction term $(g/n)\,\Phi^4(\bm{x})$ with $g/n=\tilde{u}(\bm{0})=\lambda_0m^{-2\sigma}$. The same applies to the equation of state. Thus $\gamma_\infty$ and $\nu_\infty$ have the given  ${n=\infty}$ values. However, taking the limits ${n\to\infty}$ in Eqs.~(27)--(33) of Ref.~\cite{SDN15} yields results  that are inconsistent with the exact ${n=\infty}$ values of these critical exponents, such as $\nu_{\text{SDN},\infty}= (2+\epsilon_\sigma)/4+O(\epsilon_\sigma^2)$, the only exception being $\eta_{\text{SDN},\infty}=0$. It must be emphasized that the large-$n$ limit is mathematically fully controlled. The exact large-$n$ results just mentioned therefore provide mathematically rigorous evidence against the $O(\epsilon_\sigma)$ series expansions of the critical exponents given in Ref.~\cite{SDN15}.

To summarize, the results of \cite{SDN15} are incorrect and  should not be used in comparisons with experimental results.

I close with some remarks meant  to prevent trivial misunderstandings when considering a situation where $a\ll\xi(T)\ll\xi_m$. Above, we were always concerned with the behavior in the scaling limit $\xi_m/\xi(T)\to 0$, which one must investigate to determine the universality class of the model that describes its asymptotic critical behavior for any $\xi_m<\infty$ as $T\to T_c$.
One might, of course, also be interested in the behavior in a regime of distances $|\bm{x}|$ and temperatures where both $|\bm{x}|$ and $\xi(T)$ are large and satisfy $a\ll |\bm{x}|$, $\xi(T)\ll\xi_m$. A natural way to do this is to take the limit $m\to 0$ ($\xi_m\to\infty$), i.e., to choose the algebraically decaying potential $u_{m=0}(\bm{x})\propto  |\bm{x}|^{-(d-2\sigma)}$. The authors of Ref.~\cite{SDN15} do not do this, maintaining a nonzero $m$. In the just specified regime, the ratio $|\bm{x}|/\xi(T)$ can take any value $\in (0,\infty)$; the lower and upper bounds  correspond to the limiting behaviors for   $|\bm{x}|\ll\xi(T)$ and $|\bm{x}|\gg\xi(T)$, respectively.

When ${m=0}$, power counting indeed yields the marginal dimension $d_\sigma =4+2\sigma$ for the quartic interaction term $\propto u_{m=0}$ of $\mathcal{H}_{m=0}$. In the equivalent description via $\mathcal{H}'_{m=0}$, one obtains  a quadratic $\bm{k}$-space contribution $k^{2\sigma}\psi_{\bm{k}}\psi_{-\bm{k}}/2$  without a mass term $m^2_\psi\,\psi^2(\bm{x})/2$. Power counting  again yields  the marginal dimension $d_\sigma=4+2\sigma$, which is $<4$ when $\sigma<0$. Needless to say that one should not be surprised that power counting yields different upper dimensions for the cases $m>0$ and $m=0$. Universality classes are well known to depend on gross features of the interaction. As expounded above, the  potential $u(\bm{x})$ has different gross features depending on whether $m>0$ or  $m=0$ since it is short-ranged or long-ranged on the scale of the correlation length, respectively.

It may be tempting to  perform a RG calculation for this ${m=0}$ case. (Needless to say, that the critical exponents of a corresponding new universality class should be independent of microscopic details.) However, one should also be aware that a  proper analysis  requires answers to further questions. Clearly, a better justification of the model and/or possible improvements are needed. Aside from  clarifying the origin of the long-range interaction $u_{m=0}(\bm{x})$, one should include all generically possible interactions in the starting Hamiltonian $\mathcal{H}_{m=0}$ except those that can be trusted to be irrelevant in the RG sense. In the equivalent reformulation of the model via $\mathcal{H}'_{m=0}$, the potential $u_{m=0}(\bm{x})$ requires the absence of a mass term for the $\psi$-field. This, and hence the origin of $u_{m=0}(\bm{x})$, ought to be explained. 

The situation must not be confused with that of an $n$-component $\Phi^4$ model with $\Phi$-$\Phi$ (``spin-spin'')  interactions decaying as $|x|^{-(d+\kappa)}$ with $\kappa<2$ studied in \cite{FMN72} (Ref.~[9] of \cite{SDN15}).  Unlike the present model with ${m=0}$,  the model considered in \cite{FMN72} involves an algebraically decaying  $\Phi$-$\Phi$ (``spin-spin'') interaction rather than one with the  long-ranged potential $u_{m=0}(\bm{x})$ between the energy-density operators $\bm{\Phi}^2(\bm{x})$ and $\bm{\Phi}^2(\bm{x}')$.

\begin{acknowledgments}
Helpful correspondence with H.~K. Janssen and a critical reading of the manuscript by S.B.\ Rutkevich are gratefully acknowledged.
\end{acknowledgments}

\end{document}